\documentclass[prb,reprint,superscriptaddress,letterpaper]{revtex4-1}
\usepackage{siunitx}
\DeclareSIUnit\BohrMagneton{\ensuremath{\mu_{\text{B}}}}
\DeclareSIUnit\formulaunit{f.u.}
\DeclareSIUnit\atomicunit{a.u.}
\DeclareSIUnit\arbunit{arb.unit}
\DeclareSIUnit\torr{Torr}
\DeclareSIUnit\rpm{RPM}
\usepackage{graphicx}

\newcommand{\mrx}{\mbox{Mn$_2$Ru$_x$Ga}}
\newcommand{\mrg}{\mbox{MRG}}
\newcommand{\exx}{\mbox{$\tilde{\varepsilon}_{xx}$}}
\newcommand{\exy}{\mbox{$\tilde{\varepsilon}_{xy}$}}

\newcommand{\tcmp}{T_{\text{comp}}}
\newcommand{\alox}{\mbox{AlO$_x$}}
\newcommand{\eps}[1]{\mbox{$\tilde{\varepsilon}_{#1}$}}

\begin{document}
\title{Magneto-optic Kerr effect in a spin-polarized zero-moment ferrimagnet}
\author{Karsten Fleischer}
\affiliation{School of Physics, Trinity College Dublin}
\affiliation{Dublin City University}
\author{Naganivetha Thiyagarajah}
\author{Yong-Chang Lau}
\author{Davide Betto}
\author{Kiril Borisov}
\affiliation{CRANN and School of Physics, Trinity College Dublin}
\author{Christopher C.~Smith}
\author{Igor V.~Shvets}
\affiliation{School of Physics, Trinity College Dublin}
\author{J.M.D.~Coey}
\author{Karsten Rode}
\affiliation{CRANN and School of Physics, Trinity College Dublin}

\date{\today}

\begin{abstract}
\label{sec:abstract}
The magneto-optical Kerr effect (MOKE) is often assumed to be proportional to
the magnetisation of a magnetically ordered metallic sample; in metallic
ferrimagnets with chemically distinct sublattices, such as rare-earth
transition-metal alloys, it depends on the difference between the sublattice
contributions. Here we show that in a highly spin polarized, fully compensated
ferrimagnet, where the sublattices are chemically similar, MOKE is observed
even when the net moment is negligible. We analyse the spectral ellipsometry and
MOKE of \mrx, and show that this behaviour is due to a highly
spin-polarized conduction band dominated by one of the two manganese
sublattices which creates helicity-dependent reflectivity determined by a broad
Drude tail. Our findings open new prospects for studying spin dynamics in the
infra-red.
\end{abstract}

\maketitle

\section{Introduction}
\label{sec:intro}
Spin valves, where angular momentum can be transferred from a
reference layer to a free layer, are expected to form the basis of future memory
devices and oscillators operating in the far infra-red frequency
domain.\cite{BettoAIPAdvances2016}
The key materials properties necessary to achieve these frequencies are a high
Fermi-level spin polarization $P$; low net magnetization $M_s$ and a high
effective magnetic anisotropy field $\mu_0H_{\text{k}}$.
A compensated ferrimagnetic half metal where negligible magnetization can
be obtained, coupled with full $\left( P=\SI{100}{\percent} \right)$ spin
polarization, would be an ideal choice.\cite{HakimiJAP2013}

Such a material was suggested by \citet{vanLeuken1995} in \citeyear{vanLeuken1995}, but the first
experimental evidence of one was only provided by the work of
\citet{KurtPRL2014} in \citeyear{KurtPRL2014}.
We have recently shown that the material, \mrx\ (MRG), shows negligible
$M_s$ over a wide temperature and thermal treatment range;\cite{Betto2015,7929426} high spin
polarization\cite{KurtPRL2014} as reflected in the anomalous Hall
effect;\cite{Thiyagarajah2015} produces sizeable magnetoresistive effects
in spin electronic structures\cite{doi:10.1063/1.5001172} such as magnetic tunnel
junctions;\cite{BorisovAPL2016} and proposed an \emph{ab initio} model
reconciling the experimental observations with density functional theory (DFT) band
structure calculations.\cite{ZicPRBRapid2016}

Time-resolved magneto-optical Kerr/Faraday effect is particularly suited to
determine the Larmor frequency of thin-film samples, especially when the 
resonance frequency exceeds the experimental range of cavity-
or network-analyser-based techniques. The unique properties of \mrg\ 
indicate that the material may be a candidate for low-power all-optical
switching as the magnetic compensation temperature $\tcmp$ can be tuned to lie just
above room temperature,\cite{VahaplarPRB2012} but these measurements on new
materials are complicated by the lack of an accepted,
wavelength-dependent optical model.  

Ferrimagnets with two chemically distinct sublattices, notably the rare-earth
transition metal alloys have been investigated by spectroscopic
MOKE.\cite{PhysRevLett.110.107205,doi:10.1063/1.4966183} The spectral response of the $4f$ and $3d$ elements allow
the temperature dependence of the Kerr effect of each sublattice to be
determined. The new aspect we present here is that in a ferrimagnet with chemically
identical, but structurally inequivalent, sublattices it is possible to track the
spin polarization associated with only one of the sublattices.

We first show that even at magnetic compensation, \mrg\ exhibits a clear Kerr
rotation ($\theta_K$). Spectroscopic ellipsometry, in combination with
the spectroscopic polar MOKE\cite{Kerr1877} allow us to determine the diagonal
(\exx) and off-diagonal (\exy) dielectric tensor components as a function of Ru
concentration $x$. Polar MOKE is often used simply to measure hysteresis loops
using monochromatic light, and the spectroscopic variant provides more specific
information on the sites responsible for the magnetic response. Our
measurements show a clear correlation between the infrared Kerr rotation and
the electrically-measured Hall angle, providing contactless, optical
measurements of this quantity.  The experimentally determined, optical model
developed here will facilitate measurements in the time- and frequency-domains,
that can advance the development of spin electronic devices operating in the
far infra-red frequency range.

\section{Sample details}
\label{sec:samples}
\begin{figure}
  \includegraphics[width=0.8\columnwidth]{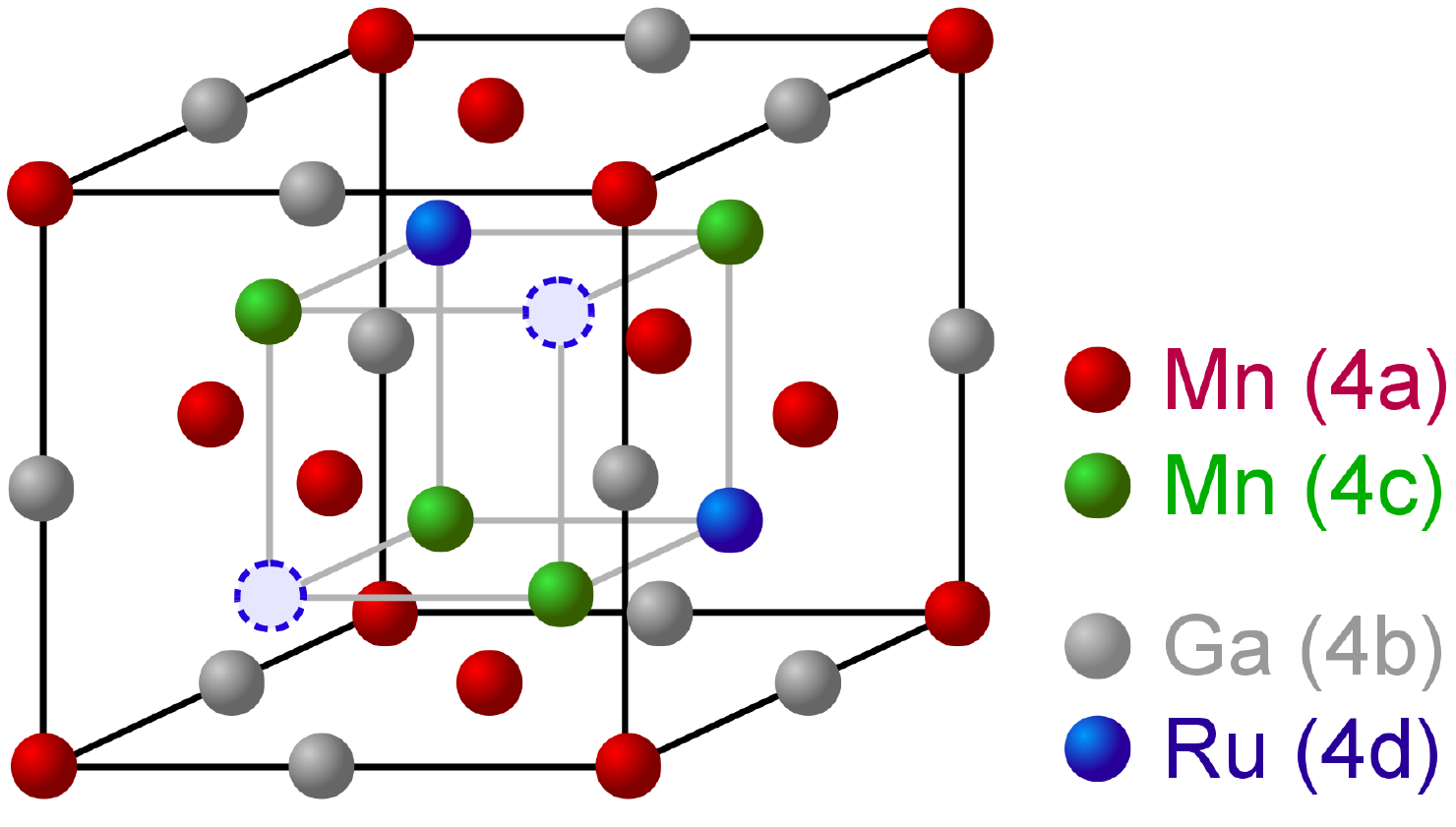}
\caption{Schematic of the \mrg\ crystal structure. There are three
  filled fcc sublattices (red: Mn-site ($4a$), grey: Ga-site ($4b$), green:
  Mn-site ($4c$)). Ruthenium occupies the remaining fcc sublattice (blue:
  Ru-site ($4d$)).  The two Mn sites are non equivalent in the structure as the
$4a$-site has only Ga as nearest neighbour, while the $4c$-site Ru, or
vacancies (dashed circles) depending on the Ru content
$x$.}\label{Fig-Structure}
\end{figure}
The thin-film samples studied here were grown on MgO $(001)$ substrates by dc
magnetron sputtering at \SI{250}{\degreeCelsius} substrate temperature in a
'Shamrock' system with a base pressure of \SI{2e-8}{\torr}. The films were
co-sputtered from Ru and Mn$_2$Ga targets. The Ru concentration
was controlled by varying the Mn$_2$Ga target plasma power while keeping that
of the Ru fixed. The samples were then capped with a protective
\SI{2}{\nano\metre} layer of amorphous aluminium oxide. The crystal structure, lattice parameters
and strain of the films were characterised by $2\theta - \theta$ X-ray
diffraction and reciprocal space mapping using a Bruker D8 Discover instrument, with a
monochromatic copper K$_{\alpha 1}$ source. We found that the in-plane lattice
constant $a$ is \SI{0.5956}{\nano\metre}
$=\sqrt{2}a_{\textrm{MgO}}$ for all samples, indicating that \mrg\ grows epitaxially
on MgO with its basal plane rotated by \SI{45}{\degree} with respect to the
substrate. The out-of-plane parameter $c$ varies between
\SIlist{0.598;0.618}{\nano\metre}, depending on Ru concentration and 
film thickness ($\sim$ \SIrange{26}{81}{\nano\metre}). This confirms a
slight, substrate-induced, tetragonal distortion of the cubic structure. An illustration of the 
crystal is given in \figurename~\ref{Fig-Structure}.

In order to determine the Ru concentration, we deposited four \mrg\ samples
with varying Mn$_2$Ga target power and a pure Ru film. The density and
thickness were then measured using X-ray reflectivity (XRR).
Based on the measured density and lattice parameters of these five control
samples, we established a relation between X-ray density and Ru
concentration $x$, against which all the samples were calibrated. The transport
properties of the thin films were measured on
non-patterned films in a Quantum Design physical properties measurement system
(PPMS) at temperatures ranging from \SIrange{10}{400}{\kelvin} in an applied
magnetic field, $\mu_0H$, of up to \SI{12}{\tesla}.

\section{Experimental details}
\label{sec:experiments}
Both ellipsometry and magneto-optical Kerr spectroscopy were used
to determine the optical and magneto-optical properties of the films. 
\label{ssec:exp_moke} Spectroscopic MOKE was recorded at
near-normal incidence with a in-house built reflectance anisotropy spectrometer,
based on the design of Aspnes.\cite{PhysRevLett.54.1956} Light from a Xe lamp
passes through a Rochon Mg$_2$F polariser and is reflected from the sample. The
beam then passes through a photo-elastic modulator (PEM), an analysing
polariser and a monochromator before finally reaching a diode detector system.
All measurements were performed from \SIrange{0.35}{5}{\electronvolt}, using a
Bentham TMc300 triple-grating monochromator and three individual photo-diodes
(InAs, InGaAs, Si). The procedure for extracting the MOKE rotation has been
discussed elsewhere.\cite{Fleischer2014-MOKE,Herrman2006,Cunniffe2010}  All
samples where measured at room temperature in remanence (no external field).
Samples were magnetised before these measurements in fields of up to \SI{12}{\tesla}
at room temperature. Selected sample have also been magnetised at 200\,K.

To correct for any set-up or non-magnetic anisotropy, we recorded spectra for
two orthogonal in-plane directions as well as for samples magnetized in
opposite directions. The spectra obtained by reversal of $\vec{M}$, were
strictly equal to those derived from a rotation of the sample of
\SI{90}{\degree} around the optical axis.

Due to larger errors in the Kerr-ellipticity measurement in a PEM setup without
\emph{in-situ} magnetisation reversal, only the Kerr rotation ($\theta_K$) will
be discussed.   

\label{ssec:exp_ellipsometry}
A Sopra GESP 5 spectroscopic ellipsometer was used over an energy range from
\SIrange{1.5}{5}{\electronvolt} with incidence angles of \SIlist{70; 75;
80}{\degree}. Both, the thin film dielectric function (\exx, \exy), as was
derived from the raw measurements by minimising the deviation between measured
ellipsometric data ($\tan\Psi, \cos\Delta$) over the whole spectral range, as
well as all three angles. Full transfer matrix calculations of the
air/AlO$_x$/MRG/MgO layer stack have to be evaluated, as in contrast to simple
bulk samples no analytical expression for the relationship between measured
$\tan\Psi, \cos\Delta$, and $\theta_K$ exists. The spectral shape of $\tilde{\varepsilon}$ was
simulated by a standard Drude-like component and three harmonic oscillators
(Drude-Lorentz oscillator). To reduce uncertainties in \exx, only the \mrg\ layer was
modelled. The sample thickness and interface roughness, as measured by XRR was
used as input parameters together with the known bulk dielectric functions of
the MgO substrate and the \alox\ capping layer.

In thin film structures, the $\theta_K$ spectra depend not only on
the off-diagonal dielectric tensor components $\eps{xy}=iQm_z\eps{xx}$ of the
magnetic material but also on the overall reflectance of the stack. Knowing the
diagonal components \exx\ from the ellipsometric model, the measured $\theta_K$
spectral signature can then also be modelled, and hence $\eps{xy}$ determined using
the same methodology without skewed results caused by variations in \exx\ and sample thickness/roughness and
modulation of the overall reflectivity.\cite{Kudryavtsev2002,Halilov1991,Zak1991}

For all optical models, the \mrg\ and \alox\ thicknesses determined by
XRR were treated as fixed parameters. All interfaces have been treated as sharp,
while the dielectric function of the \alox\ coating has been modelled by a
Bruggeman effective medium of crystalline Al$_2$O$_3$ and air with a fill
factor set to the density ratio of the XRR fitted capping layer
and crystalline Al$_2$O$_3$. We treat the \mrg\ to be perfectly cubic, as the deviation from
cubic symmetry due to epitaxial strain of $\sim\SI{1}{\percent}$\cite{Kurt2014} is well 
below the sensitivity of our instrument.

\section{MOKE: Direct Measurements}
\label{sec:MOKE}
\begin{figure}
  \centering
  \includegraphics[width=\columnwidth]{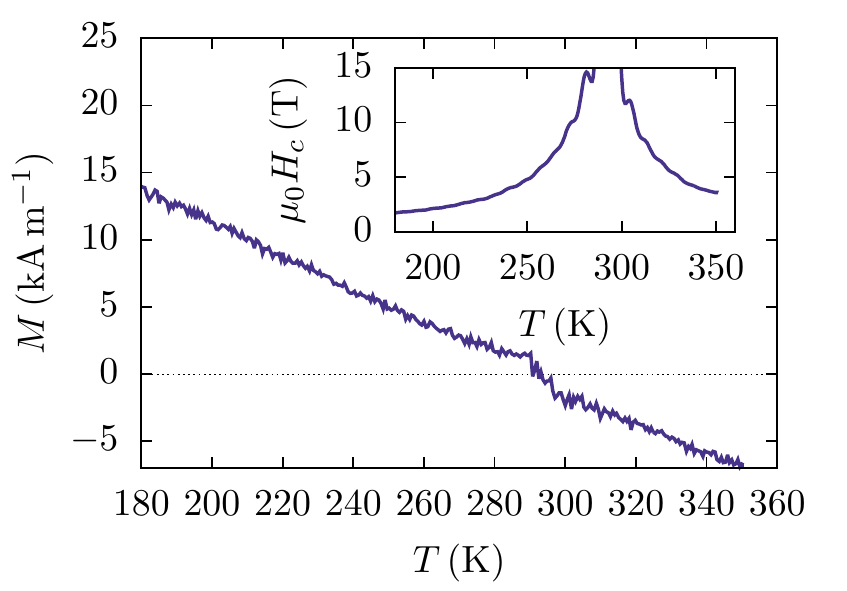}
  \caption{SQUID magnetometry of sample with $\tcmp$ close to RT. The net
    magnetisation of the sample was measured at remnance on warming after
    saturation at low temperature. In the inset we show an estimate of the
    coercive field, $\mu_0 H_c$, as a function of temperature.
  Around room-temperature, even $ \SI{15}{\tesla}$ is insufficient to saturate
the magnetisation.}
  \label{fig:MT}
\end{figure}
Due to the different crystallographic sites occupied by the two magnetic
sublattices in ferrimagnets, the temperature dependence of the
magnetisation of the two sublattices differ. Generally, perfect compensation
and zero net moment occurs only at a specific temperature
($\tcmp$) where the magnitudes of the two magnetic sublattices are equal but
opposite in sign. For \mrg\ this temperature depends on both Ru content
$x$ and the biaxial substrate-induced strain.

Upon crossing $\tcmp$ in zero applied field the net
magnetisation changes sign whereas the spin-dependent DOS remains
unchanged because the individual sublattices have not changed direction. We
therefore expect that the sign of any property that depends on the
spin-resolved DOS such as MOKE, tunnel magneto-resistance
(TMR) and anomalous Hall effect, will
exhibit opposite signs depending on whether the sample was magnetised below or above
$\tcmp$.\cite{BorisovAPL2016,Thiyagarajah2015}

In \figurename~\ref{fig:MT} we show the temperature dependence of the net
magnetic moment of an \mrg\ sample with $\tcmp \sim \SI{295}{\kelvin}$. The
inset gives an estimate of the coercive field as a function of temperature.
Close to $\tcmp$, the anisotropy field $\mu_0 H_k$ diverges, and as the
coercive field is directly related to the anisotropy field, the sample cannot
be magnetically saturated in this temperature range.

\begin{figure}
  \centering
  \includegraphics[width=\columnwidth]{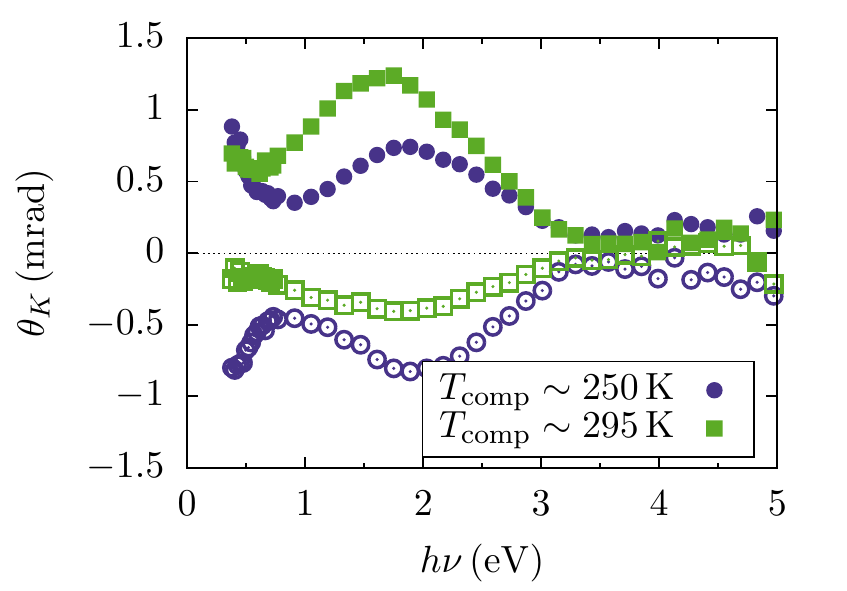}
  \caption{Room-temperature Kerr rotation as a function of photon energy for
    samples with $\tcmp \sim \SIlist{295;250}{\kelvin}$, respectively. The
    samples were first magnetised at RT in a field of $\mu_0 H =
    \SI{2}{\tesla}$ (above $\tcmp$, open symbols).  Subsequently, the samples
    were remagnetised, this time well below $\tcmp$ ($T_{\text{sat}}
      = \SI{200}{\kelvin}$, filled
    symbols). Although all measurements are done at RT, the sign
    of the MOKE signal is reversed depending on whether magnetisation is carried out
    above or below $\tcmp$.  When $\tcmp \sim \SI{295}{\kelvin}$, full saturation
    is only achieved when magnetised well below $\tcmp$. }
  \label{fig:moke-tcmp}
\end{figure}
We now compare MOKE spectra of two samples with $\tcmp$ of
\SIlist{295;250}{\kelvin}, respectively. The data are shown in
\figurename~\ref{fig:moke-tcmp}. When magnetised at RT, \emph{i.e.} above
$\tcmp$, both samples exhibit clear MOKE \emph{negative} in sign. The sample
with $\tcmp$ closest to RT appears to have a reduced magnitude. We then
magnetise both samples well below their respective $\tcmp$, $T_{\text{sat}} =
\SI{200}{\kelvin}$, and subsequently warm to RT in zero applied field. This
procedure ensures that while the net magnetic moment changes sign when crossing
$\tcmp$, the spin-dependent density of states (DOS) remains unchanged. We then
record MOKE spectra at RT. As can be seen from \figurename~\ref{fig:moke-tcmp},
both samples exhibit a change of sign as expected, and crucially, the sample
with $\tcmp \sim \SI{295}{\kelvin}$ and thus $\sim$ zero net moment now
displays as strong or stronger MOKE signal as the sample measured further away
from $\tcmp$. This set of measurements proves that MOKE in \mrg\
depends on the spin-dependent DOS, and not the net magnetic moment; and
furthermore even when the net magnetic moment is zero, the MOKE signal
is clearly observed.

We note here that disproportionally between the remnant magnetization
and the polar MOKE measured during a hysteresis loop have been observed
earlier.\cite{Osgood1997,Tillmanns2006} This was attributed coupling, \emph{during
switching}, of the
in-plane magnetization and the polar MOKE through higher order terms. It was only observed
in highly anisotropic exchange-biased materials and only during switching.
The samples measured here do not exhibit uniaxial in-plane anisotropy, and are all measured
at remanence.

In order to pinpoint the origin of MOKE in \mrg , we recorded angle
and energy dependent ellipsometry, which in combination with the measured Kerr rotation
allow us to determine all matrix elements of the dielectric tensor.

\section{Optical Model}
\label{sec:optical_model}
We have shown that MOKE is not determined by the net magnetic moment, but depends on the spin 
dependent DOS in \mrg.
MOKE measurements alone are insufficient to determine \exx\ and \exy.
We therefore first use spectroscopic
ellipsometry to determine \exx\ and subsequently develop the optical model
including the off-diagonal tensor elements \exy\ for \mrg.

\begin{figure}
  \centering
  \includegraphics[width=\columnwidth]{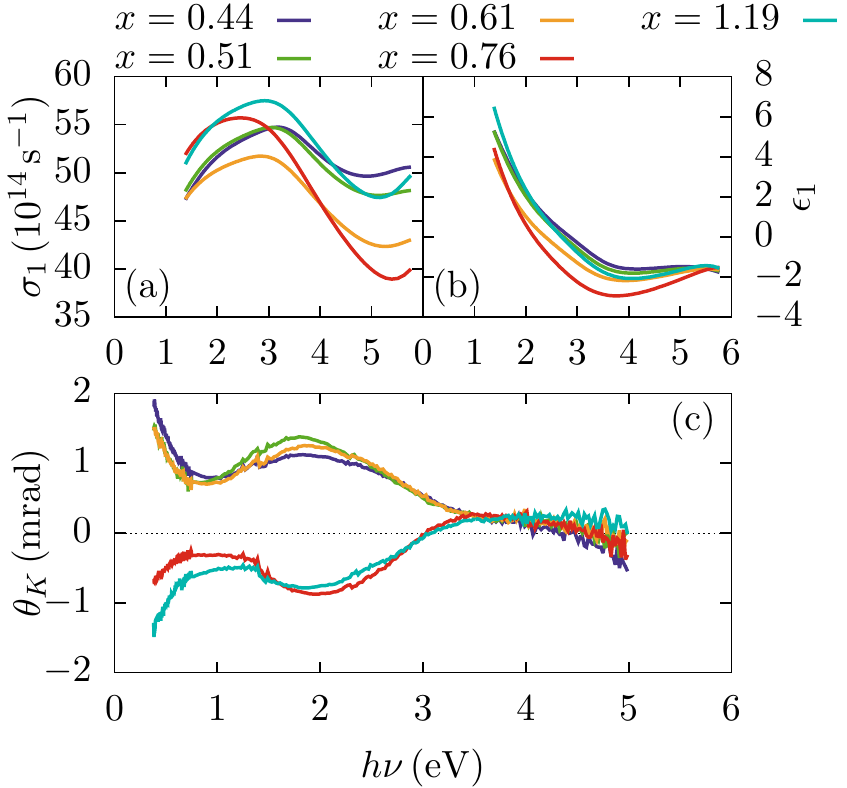}
  \caption{(a) and (b): Imaginary and real components of the diagonal tensor
    elements \exx\ of the dielectric function of \mrg\ as a function of photon energy
    and Ru concentration $x$. The imaginary part is shown in terms of optical
    conductivity ($\sigma_1 = \varepsilon_2 \omega / 4 \pi$)\cite{Johnson1974}, to
    highlight the small differences between samples.The behaviour is dominated
    by the Drude-like metallic response, with smaller deviations caused by
    inter-band transitions. (c) MOKE spectra for the same series of samples
    used to infer the off-diagonal \exy\ tensor elements. }
\label{fig-eps}
\end{figure}
In \figurename~\ref{fig-eps}  we show the dielectric function \exx\ as extracted from
least square fits of the measured $\tan\Psi$ and $\cos\Delta$ for a set of samples with 
varying Ru concentration $x$. Our optical model, used for both \exx\ and  \exy\, is based on 
the following considerations:

For metallic materials such as \mrg\ the dielectric response is usually
dominated by a strong Drude-like tail caused by free electron absorption, also
known as \emph{intra-band} transitions.\cite{YuCardona} This contribution is
the reason why the imaginary part of \exx\ of \mrg\ does not tend towards zero
($\sigma \sim \SI{45E14}{\per\second}$ at $h \nu = \SI{1}{\electronvolt}$) in
the infra-red region.  Furthermore, there are absorption features in the
visible (VIS - peaks $\sim \SI{2.3}{\electronvolt}$) which we attribute to
interband transitions on both Mn sites, ultra-violet (UV $\sim
\SI{3.2}{\electronvolt}$) due to Ga, and a deep ultra-violet contribution (DUV $>
\SI{5.5}{\electronvolt}$) necessary to account for the line shape but that cannot
be assigned to a specific element. The assignment of the VIS structure to
Mn interband transitions is motivated by several experimental and theoretical
findings, including similar spectral features in bulk Mn,\cite{Johnson1974}
a strong peak in the unoccupied density of states in \textit{ab-initio}
calculations for bulk Mn,\cite{Hobbs2003} and finally the similarity in the shift of the 
energetic position of the VIS peak with increasing Ru concentration, following
the same trend as that seen in site-selective XAS/XMCD measurements on \mrx.\cite{Betto2015}

To minimize the number of free parameters, we therefore describe the diagonal
part of the dielectric function via one Drude-like oscillator at $\omega =
\SI{0}{\electronvolt}$, and three broad harmonic oscillators ($\Gamma >
\SI{6}{\electronvolt}$) in the VIS, UV and DUV ranges, respectively.

\begin{figure}
  \centering
  \includegraphics[width=\columnwidth]{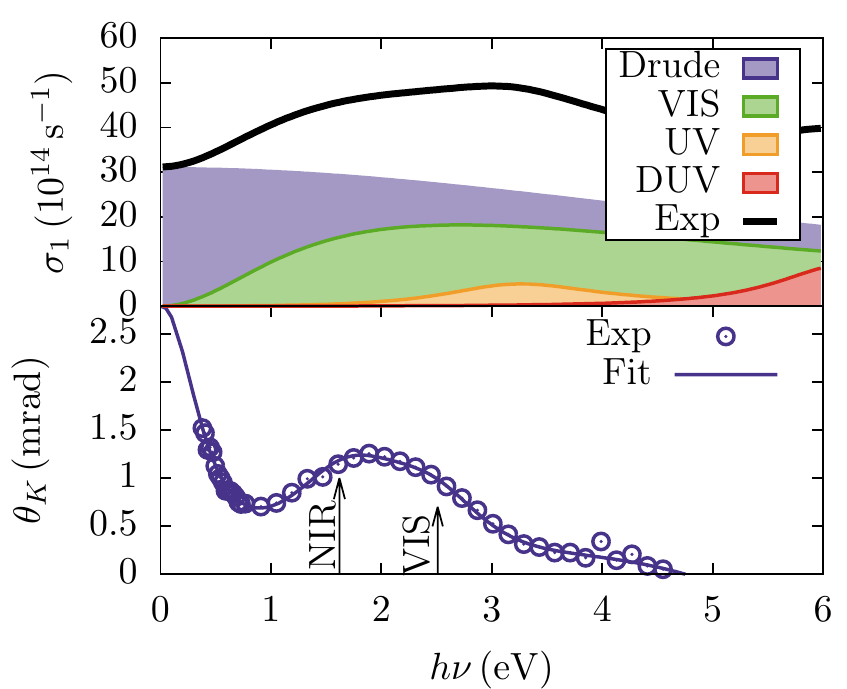}
  \caption{Complete optical model and experimental data for an \mrg\ sample
    with $x = 0.61$. (Top) Experimentally observed optical conductance (line) and the fitted
contributions from the four harmonic oscillators (areas) described in the text.
(Bottom) Experimental Kerr rotation (points) and  fit to the data
  (line) based on the optical model described in the main text.}
\label{fig-simus}
\end{figure}

The Kerr spectra show strong features in the IR and the VIS regions,
associated with the Drude-tail and the Mn, respectively, but no separate features
in the UV and DUV. As the magnetism in \mrg\ is dominated by Mn in two
antiferromagnetically coupled sub-lattices, \exy\ can be described in more
detail. We associate the single, broad, VIS oscillator observed in ellipsometry
at $h \nu \sim \SI{2.3}{\electronvolt}$ with two individual components in the Kerr spectra (NIR and VIS in
\figurename~\ref{fig-simus}, bottom panel), which we assign to the spin-dependent
transitions on the Mn-$4c$ and Mn-$4a$ site.  In agreement with
our assignment above of the UV and DUV transitions to non-magnetic components,
no separate UV and DUV oscillators are required to describe \exy.

This assignment is motivated by our earlier X-ray absorption
measurements\cite{Betto2015} where we find that the final states of the
$2p\rightarrow 3d$ transitions of Mn in the two sites are only separated by
approximately \SI{1}{\electronvolt}, and no circular magnetic dichroism on the
Ga $L_{2,3}$ edges nor on the Ru $M_{4,5}$ edges, as well as the dielectric
functions of Mn\cite{Johnson1974} and Ga\cite{Hunderi1974}.

In \figurename\ \ref{fig-simus} we show an example of a fit to the model
described above. Having established the diagonal tensor elements \exx\, we then
model the off-diagonal terms which we plot as the expected Kerr rotation
$\theta_K$ (the real part of the complex Kerr rotation). The agreement
between the experimental data and the optical model is excellent, indicating that
our model is well justified as well as highlighting the
predominant role of Mn in \mrg\ magnetisation. The (extrapolated) steep
increase in MOKE in the IR and towards DC ($h \nu \rightarrow
\SI{0}{\electronvolt}$) is clearly an effect of the
highly spin-polarised conduction band, as reflected by the Drude
contribution to both diagonal and off-diagonal tensor elements. 

Our modelling allows us to study in finer detail the behaviour of the spectral
components of the ellipsometry and MOKE as a function of Ru concentration $x$
(\figurename~\ref{fig-eps}).

When the Ru concentration $x$ changes from \numrange{0.44}{1.19}, the
oscillator associated with Ga (UV) changes position only slightly from
\SIrange{3.4}{3.2}{\electronvolt} whereas the oscillator associated with Mn
(VIS), exhibits a stronger trend and moves from
\SIrange{3.0}{2.2}{\electronvolt} with increasing $x$. With increasing Ru
content, the electronic occupation of Mn is increasing linearly, wheres that of
Ga remains constant. The (small) changes associated with Ga are thus due to
increasing tetragonality of the crystal structure while those associated with
Mn correlate directly with the band filling. As described above, the MOKE in
\mrg\ is dominated by the contributions of Mn, both through the spin
polarisation of the conduction band as well as single-ion transitions.
The two oscillators used to describe these transitions in MOKE can therefore be
associated with Mn in the $4c$ position ($\sim \SI{1.7}{\electronvolt}$ - NIR)
and in the $4a$ position ($\sim \SI{2.5}{\electronvolt}$ - VIS). Due to the
half-metallic nature of the sample, when Ru content is increased, the
electronic occupation of Mn in the $4c$ position increases leading to a blue
shift (from \SIrange{1.4}{1.8}{\electronvolt}) of the oscillator associated
with this position. Simultaneously, adding Ru decreases the
spin-gap\cite{ZicPRBRapid2016} thereby inducing a red shift (from
\SIrange{2.7}{2.4}{\electronvolt}) of the oscillator associated with Mn in the
$4a$ position. More details on the optical model parameter as function of 
Ru concentration $x$ are found in the supplemental information.

\begin{figure}
  \centering
  \includegraphics[width=\columnwidth]{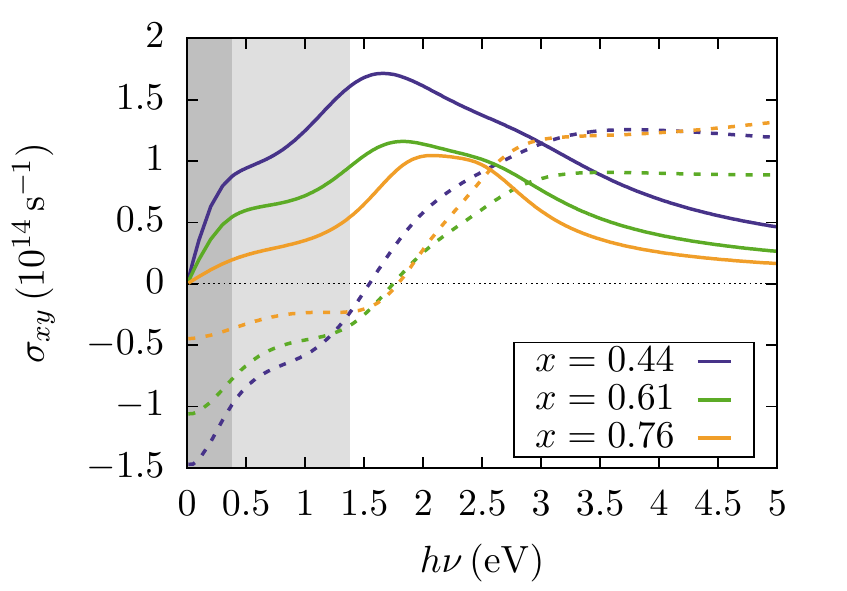}
  \caption{$\sigma_{xy}$ extracted from the optical model for three different Ru
    concentrations $x$. The real and imaginary parts are plotted in solid and
broken lines, respectively. In the light and dark grey areas, the curves are
extrapolated from the experimentally available range in ellipsometry and MOKE,
respectively. Please not that Im($\sigma_{xy}$) is only determined up to a
potential integrational constant}
  \label{fig:eps-sim}
\end{figure}
In \figurename\ \ref{fig:eps-sim} we show the off-diagonal component of the
dielectric tensor as a function of photon energy, extracted from the fit of the
experimental ellipsometry and MOKE data, in terms of $\sigma_{xy}$. Unlike
the raw MOKE spectra, differences between different compositions are now clearly visible, and
highlight, independently of our discussion focussed on \mrg, the need to
employ robust optical modelling when reporting MOKE data on thin film samples of 
even slightly varying thickness.

We finally turn to a comparison of the DC limit of the magneto-optical
properties with the anomalous Hall angle measured by standard electronic
magnetotransport.

\begin{figure}
  \centering
  \includegraphics[width=\columnwidth]{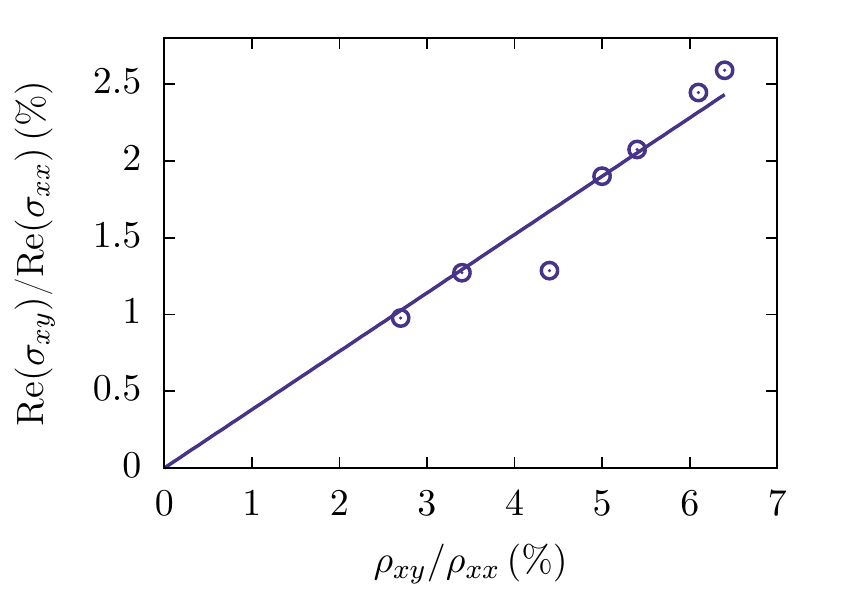}
  \caption{Ratio of the real part of the off-diagonal and on-diagonal terms of
    the dielectric tensor extrapolated to \SI{0.4}{\electronvolt} plotted as a function of the electrically measured
anomalous Hall angle. The straight line is a fit to the data, and
provides a guide to the eye. As expected, the DC limit of the magneto-optical
characterisation agrees very well with the values obtained through electronic
transport measurements.}
  \label{fig-hallangle}
\end{figure}
As outlined above, our optical model describes the behaviour of \mrg\ fully in
the spectral region measured, and we therefore expect that extrapolation
outside this region should accurately predict the behaviour at different photon
energies. The lower limit is of course DC where optical measurements are not
possible, but where we can record magnetotransport, in particular anomalous
Hall effect.\cite{Thiyagarajah2015} The ratio $\rho_{xy}/\rho_{xx}$ is the
Hall angle, the DC limit of the ratio Re$(\sigma_{xy})$/Re$(\sigma_{xx})$. In
\figurename~\ref{fig-hallangle} we plot the ratio determined from the
extrapolation of our optical model as a function of the Hall angle determined 
by magnetotransport measurements as described in \ref{sec:samples}.
The ratio was evaluated at $h \nu = \SI{0.4}{\electronvolt}$ as we only need to
extrapolate $\sigma_{xx}$ and not $\sigma_{xy}$ at this energy.  As expected we observe a strong correlation
between the two measurements, most certainly because the two effects share a
common origin; the high spin polarisation of the conduction band. The
deviations from a perfectly proportional behaviour is due to small differences in the
scattering matrices that have a big influence in transport measurements whereas
they do not influence the values obtained by optical means. We also emphasize
that whereas electronic transport probes the entire thickness of a metallic film, MOKE
and ellipsometry only probe the
skin-depth, which varies with photon energy,
found here to be \SI{5}{\nano\metre} at \SI{2}{\electronvolt} and \SI{40}{\nano\metre}
at \SI{0.4}{\electronvolt}.
\section{Conclusion}
\label{sec:conclusion}
We have shown that MOKE in zero-moment half metals persists, even when the net
magnetic moment crosses zero, and that its origin is dominated by the
highly spin-polarised conduction band associated in \mrg\ with Mn in the
$4c$ site.\cite{Betto2015} The sign of the MOKE depends on whether
the sample was magnetised above or below its compensation temperature, as the
process of magnetisation sets the direction of the axis of quantization for the
spin-polarised density of states.

By combining spectroscopic ellipsometry and MOKE it was possible to construct
an optical model that identifies the optically active components of \mrg , and we
use this to infer the full parameterized dielectric tensor as a function of
photon energy in the range \SIrange{0}{5}{\electronvolt}. The model is 
deliberately constructed to minimize the number of free parameters, yet it captures
with reasonable accuracy both the nonmagnetic and the magnetic contributions
for
\mrg\ in the spectral range measured. Based on the model, we
compare the results obtained by optical means with the anomalous Hall angle
recorded by electronic transport measurements. Although the DC limit is far
beyond our experimentally available range, the extrapolated values of
$\sigma_{xy}$ agree remarkably well with those obtained by transport
measurements.

Differences between magnetic properties observed by anomalous Hall effect and
MOKE are often used nowadays as a fingerprint of topological electronic
transport.\cite{Matsunoe1600304}  An
extension of this study with a more in-depth comparison between the two in high and zero
applied magnetic fields, may allow an
unambiguous determination of the scattering coefficients that lead to
topologically driven spin structures in for example non-collinear ferrimagnets,
as a complete optical model is necessary to disentangle effects that depend
solely on differences in sample thickness and reflectivity from those that are
purely magnetic in nature.

Finally, this result interestingly clears a path to study the behaviour of magnetic
domains in antiferromagnets by MOKE microscopy, while the shown spectral
dependence highlights photon energies most suitable for pump-probe based time
dependent measurements.\cite{doi:10.1063/1.4958855}

\begin{acknowledgments}
  This project has received funding from the European Union's Horizon 2020
  research and innovation programme under grant agreement No 737038. KF and IVS
  acknowledges support from Science Foundation Ireland (SFI), grant number
  06/IN.1/I91. DB
  gratefully acknowledges funding from the Irish Research Council. This work was
  supported by SFI through AMBER.
\end{acknowledgments}
\bibliography{bibliography}
\end{document}